\documentclass[9pt,twocolumn,twoside]{pnas-new}
\usepackage{microtype}
\usepackage{enumitem}
\usepackage{enumitem}
\usepackage{booktabs}
\usepackage{tabularx}
\usepackage{gensymb}
\usepackage{caption}
\usepackage{xcolor}
\usepackage[utf8]{inputenc}

\templatetype{pnasresearcharticle} 

\makeatletter
\usepackage{fancyhdr}
\fancypagestyle{firststyle}{%
  \fancyhf{}
  \fancyfoot[C]{\thepage\ |\ \href{https://abbas.sitpor.org}{\texttt{abbas.sitpor.org}}}
}
\makeatother

\title{What is emergence, after all?}

\author[a, b, c, *]{Abbas K. Rizi}

\affil[a]{DTU Compute, Technical University of Denmark, Kongens Lyngby 2800, Denmark}
\affil[b]{Center for Social Data Science (SODAS), University of Copenhagen, Denmark}
\affil[c]{Department of Computer Science, School of Science, Aalto University, FI-00076, Finland}

\leadauthor{K.~Rizi} 

\correspondingauthor{\textsuperscript{*}To whom correspondence should be addressed. E-mail: \hyperlink{mailto:akari@dtu.dk}{akari@dtu.dk}}

\keywords{Emergence $|$ Reductionism $|$ Complex Systems $|$ More is Different} 

\begin{abstract}
The term emergence is increasingly used across scientific disciplines to describe phenomena that arise from interactions among a system’s components but cannot be readily inferred by examining those components in isolation. While often invoked to explain higher-level behaviors—such as flocking, synchronization, or collective intelligence—the term is frequently used without precision, sometimes giving rise to ambiguity or even mystique. In this perspective paper, we clarify the scientific meaning of emergence as a measurable and physically grounded phenomenon. Through concrete examples—such as temperature, magnetism, and herd immunity in social networks—we review how collective behavior can arise from local interactions that are constrained by global boundaries.
By refining the concept of emergence, we gain a clearer and more grounded understanding of complex systems. Our goal is to show that emergence, when properly framed, offers not mysticism but insight. \end{abstract}

\dates{This manuscript was compiled on \today}

\begin{document}

\maketitle
\ifthenelse{\boolean{shortarticle}}{\ifthenelse{\boolean{singlecolumn}}{\abscontentformatted}{\abscontent}}{}

\dropcap{E}mergence occurs where a whole exhibits properties absent from its individual parts—an idea that traces back to Aristotle’s dictum that “the whole is something besides the parts” \cite{falcon2005aristotle}. In science, the term refers to patterns, structures, or behaviors that are not straightforwardly deducible from components in isolation \cite{britannica_emergence, oxford_emergence, oconnor1994, iep_emergence, sep_emergent, kivelson2016defining}, yet are often captured by coarse, predictive descriptions that do not track every micro-detail. Emergence can be observed in the world around us: polarisation in public discourse \cite{salloum2025anatomy}, flocks of birds moving in unison, fireflies synchronizing their flashes, and insect societies coordinating their activities without central command \cite{matheny2019exotic}. It can also be studied in computer simulations, where simple local rules generate complex collective behavior—for example, Reynolds’ “boids” \cite{reynolds1987boids}, cellular automata \cite{wolfram1984cellular}, and gliders in Conway’s Game of Life \cite{barnett2023dynamical}.

\begin{figure}
    \centering
\includegraphics[width=.92\linewidth]{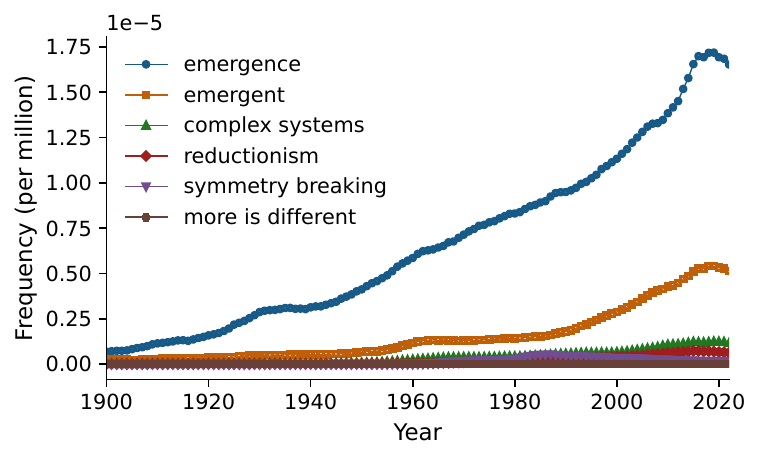}
    \caption{\textbf{Usage trajectories of key concepts in complexity discourse, 1900 – 2022.} Per-million-word frequency of six terms in the Google Books English corpus, 1900–2022. Seven-year–smoothed curves (smoothing = 3) reveal the rise of ``emergence'' and ``emergent'' compared to other famous keywords based on data downloaded via Google Ngram Viewer \cite{googleNgramEmergence}.}
    \label{fig:Ngram}
\end{figure}

Philosophical roots of the concept trace back to Mill’s distinction between additive “mechanical” effects and qualitatively new “chemical” ones \cite{mill1843logic}. Lewes later coined the term “emergent” for such cases \cite{lewes1875}, and early British emergentists \cite{alexander1920space, morgan1923emergent, broad1925mind, mclaughlin1992rise} developed it into a layered view of nature in which higher-level laws complement, and sometimes influence, lower-level processes. This tradition faded as \textit{reductionism} gained ground, driven by the intertheoretic reduction models \cite{hempel1948studies, nagel1961structure, crick1958centraldogma}. The debate revived with arguments for multiple realizability \cite{putnam1975philosophy, fodor1974special, davidson1970mental} and Anderson’s claim that “more is different” \cite{anderson1972more}, followed by work on \textit{supervenience} and causal autonomy \cite{kim2006emergence}.
In recent years, the term has evolved from describing specific transitions, such as the “emergence of metabolism” \cite{bagley1990metabolism}, to a popular buzzword, gaining greater use in scientific and non-scientific discourses \cite{holland1998emergence, anderson1972more}. Fig.~\ref{fig:Ngram} shows this trend.

Here, we review \textit{operational} perspectives on emergence—consistent with physicalism and grounded in effective theories and information-theoretic principles \cite{sethna2021statistical,jaynes1957information,presse2013principles, shannon1948}, whose thermodynamic meaning in conventional statistical mechanics is also well established \cite{presse2013principles,presse2013nonadditive,shalizitsallis}. However, there is also a rich landscape of generalized non-additive entropies and nonextensive statistical mechanics \cite{tsallis2023non, beck2003superstatistics, jizba2019maximum}, which provide alternative ways to describe emergence. We encourage interested readers to consult this literature for complementary perspectives.
In the sections that follow, we assemble the concepts required for a working account of emergence across physics, biology, and the social sciences. 
We formalize emergence through \emph{lossy compression} that preserves the information relevant for prediction and control, and we motivate \emph{Effective Theories} as the mechanism that secures autonomy and multiple realizability. 
We then treat quantitatively secure testbeds—critical phenomena and phase transitions—to show how order parameters, response functions, scaling, and data collapse locate the onset of emergence and render it mechanistic and testable. 
Next, we use dualities to show that distinct microdescriptions can yield the same macrophysics, grounding autonomy without extra causes. 
Finally, we distinguish between explanatory and ontological reduction: reduction secures cross-scale consistency, while effective theories deliver explanations and control. 
Sequenced this way, the paper justifies an economy-of-description perspective: we can explain how many emergent phenomena happen—even if not yet all—and none require supernatural causes.

\section*{Levels of Description}
Knowledge unfolds across successive levels of abstraction—from raw experience to language and higher-order concepts \cite{korzybski1994science}. Confusion between these levels distorts understanding. Emergence is everywhere, and the surprise it evokes often reflects where we choose to stop asking more questions \cite{tinbergen1963aims, kim2006emergence}. What we call “emergent” typically depends on the limits of our knowledge, tools, or perspective—in other words, on \textit{epistemology} \cite{riedl2019structures, emmeche2000levels}. 
Mixing red and green paint yields a dull brown—basic chemistry and optics explain it well enough for most, and few would call the result emergent. Yet a physicist who tracks the same blend down to quantum‐mechanical interactions would hardly find the task trivial.

Explanation is always a matter of perspective: an economist analyzing a financial crisis need not invoke atoms or molecules but instead works at the social scale relevant to the questions at hand. Every phenomenon can be analyzed at multiple descriptive levels, and what feels emergent at one level may be obvious at another \cite{riedl2019structures}. 
Even so, emergent behaviors are not merely in the eye of the beholder—they can display objective, quantifiable signatures such as information flow or causal strength \cite{ellis2018top, hoel2017map}. We can make emergence more rigorous by formalizing it through coarse-graining maps that discard detail while preserving predictive power.

\section*{Many-to-one Maps \& Coarse-graining}
\begin{figure*}
    \centering
\includegraphics[width=.965\linewidth]{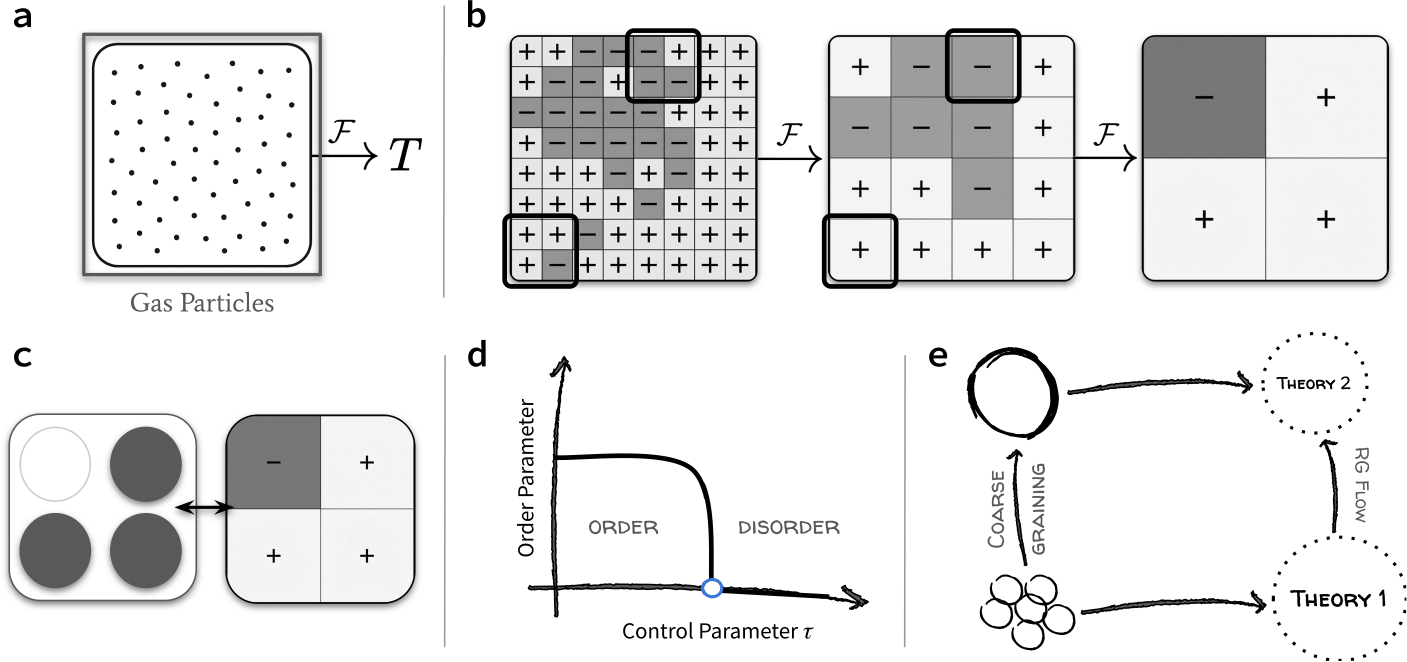}
\caption{\textbf{Coarse-graining, Universality, and RG-flow.} \textbf{(a)} Temperature as a coarse-grained variable: under local equilibrium, averaging kinetic energies defines temperature, $\mathcal{F}:(v_1,\ldots,v_N)\!\mapsto\! T$.
\textbf{(b)} Local coarse-graining by $2{\times}2$ blocks with majority-vote,
$S_I=\operatorname{sgn}(s_i+s_j+s_k+s_l)$ for $s_\ell\in\{\pm1\}$; iterating the map flows toward a uniform block-spin configuration.
\textbf{(c)} Liquid–gas criticality shares the Ising universality class. The control parameter is the reduced temperature $\tau=(T-T_c)/T_c$. On a lattice, the lattice-gas mapping $n_i\!\in\!\{0,1\}\mapsto \sigma_i=2n_i-1$ (occupied $\bullet$, empty $\circ$)  makes the density difference $\Delta\rho=\rho-\rho_c$ directly analogous to the magnetization $m$.
\textbf{(d)} Increasing temperature drives a continuous transition from an ordered phase to a disordered phase, and the transition point we see scaling behaviors \cite{sethna2021statistical, kardar2007statistical}. 
\textbf{(e)} Coarse-graining compresses high-resolution data and induces a renormalization-group (RG) flow in theory space. When the coarse-grained description is related to the microscopic one by parameter redefinitions, the model is said to be \textit{renormalizable}. A macro-law screens off micro-details but remains autonomous, meaning that macro variables aren’t just convenient summaries but “real enough” to be the subject of genuine laws and explanations—even if many different microstates realize the same macrostate.}
    \label{fig:coarse_graining_schem}
\end{figure*}

Emergence is present when there exists a many-to-one \textit{map} from a micro-level theory (more fundamental, detailed, or lower-level) to a macro-level theory (higher-level), such that the macro description remains predictive even after discarding most of the microscopic detail \cite{carroll2024emergence}.  As Fig.~\ref{fig:coarse_graining_schem}(a,b) presents, a map is a selective, structured, purpose-driven abstraction—useful precisely because it is not the whole territory it depicts \cite{korzybski1994science, wuppuluri2018map}. Every map omits, distorts, or abstracts; acknowledging those sacrifices is part of scientific hygiene \cite{wuppuluri2018map}. 

\textit{Coarse-graining} is the process of identifying which features of a system are essential for capturing its macroscopic behavior and which can be systematically discarded. Therefore, it leaves us with just enough structure and real patterns to construct an autonomous and useful theory \cite{dennett1991patterns}. This is why engineers who build bridges do not need to take a course in quantum field theory, or why airplanes can fly safely despite our incomplete understanding of quantum gravity. 
Suppose a system’s state at time $t$ is given by the vector $\mathbf{x}(t)=\bigl(x_{1}(t),\dots,x_{N}(t)\bigr)$, where a high-resolution dataset comprises $N$ time series, and each $x_i(t)$ represents the evolution of the $i$-th microscopic variable. The coarse-graining can be described as a mapping:
\begin{equation}
\mathcal{F}:\;
    \bigl(x_{1}(t_{i}),\dots,x_{N}(t_{k})\bigr)
    \;\longmapsto\;
    \bigl(X_{1}(t'_{j}),\dots,X_{N'}(t'_{\ell})\bigr),
\end{equation}
where \( N' < N \), and the coarse-grained observables \( \{X_{j}\} \) may be defined on a different (typically coarser) time grid \( \{t'_{\ell}\} \). Therefore, $\mathcal{F}$ is a \emph{lossy data compression} where only the information needed for prediction survives across scales \cite{shannon1948, jaynes1957information}. Fig.~\ref{fig:coarse_graining_schem} pictures such maps. This viewpoint offers a \textit{substrate-independent} vocabulary for linking small-scale descriptions to large-scale laws. In high-energy physics, the information relevant for prediction determines which effective theory you use at each scale \cite{maldacena1999large}. In brains and societies, the same approach helps identify mesoscales that are maximally compressed yet still predictive \cite{bialek2001predictability, tishby2000information, schreiber2000measuring, ball2010quantifying, varley2022emergence}. Panels (a) and (b) show two different sets of coarse-graining where they average out the observables. Panel (e) illustrates how, by compressing data through the right process, we arrive at a simpler theory.   
The map \(\mathcal{F}\), however, need not be simple, trivial, local, or unique.

The first step toward understanding how emergence occurs is to identify when and under what conditions it arises, as well as its immediate consequences. Useful large-scale descriptions may need entirely different degrees of freedom than the ones appearing in the small-scale theory, and they can realize symmetries in new ways. Choosing the right variables is a scientific discovery step, not a fixed recipe. The emergence of life, mind, or social structure appears puzzling not because they conflict with physicalism, but because they represent a qualitatively more complex class of emergent phenomena \cite{dedeo2018origin}. Therefore, there is no universal recipe for identifying the “right” macroscopic variables. The key practical questions here are whether there exists a finite set of macroscopic parameters that can adequately describe the phenomena of interest within a specified tolerance, and if so, which mapping \( \mathcal{F}^{\star} \) minimizes the chosen distortion measure while achieving that macro-level description. The outcome is a systematic, metric-aware approach to transitioning from micro-level to macro-level dynamics.

\section*{Reductionism \& “Theory of Everything’’}
The very notion of a map between levels of description tempts us to believe that deeper, more fundamental layers offer better explanations—an intuition reinforced by Physics’ success in \textit{reducing} systems to atoms and beyond. But how far should we go? What counts as the most fundamental level?
Physics is expressed in intrinsically redundant languages—coordinates, gauge potentials, duality frames, and so on—and those redundancies shift as we move between scales \cite{witten2018symmetry}. What we call “fundamental”—even spacetime and gauge fields—may not exist in a more microscopic theory, but rather emerge after coarse-graining. Spacetime geometry and general relativity might also be emergent, arising from quantum theory through entanglement and boundary dynamics \cite{witten2018symmetry}. This realization, however, does not make emergence a “secret sauce.” It simply shows how nature organizes complexity through scale, and “fundamental’’ is often a provisional concept. 

Each scale brings qualitatively new behaviors that demand their own inquiry. So, reducing a phenomenon to \textit{fundamental laws} does not mean we can reconstruct everything from them. \textit{Reductionism} is not wrong—it is just insufficient for understanding the universe \cite{nartallo2025reductionism}. Even a complete microscopic “Theory of Everything” would leave many of the essential rules governing higher-level systems untouched \cite{laughlin2000theory}.
Macroscopic behavior is determined by emergent parameters that are largely insensitive to the fine details of the underlying microscopic governing equation \cite{weaver1948science}. Unifying quantum mechanics and gravity would undoubtedly be a milestone in Fundamental Physics, but it would not immediately explain why financial markets crash. These are collective phenomena with different \textit{ontologies}, which cannot be derived by reduction alone \cite{laughlin2000theory}.

\section*{A New Ontology}
In Classical Mechanics, the complete state of a system is described by the positions and momenta of its particles \cite{arnol2013mathematical}. Concepts like temperature or pressure only become meaningful in the thermodynamic limit—for instance, when considering systems with $10^{23}$ particles \cite{landau2013course}. Large language models (LLMs) display a digital analogue of this behavior: they acquire new capabilities, such as multi-digit arithmetic or spatial reasoning, only after reaching a sufficient scale \cite{krakauer2025large}. In this way, temperature emerges as a qualitatively different property in the large-size limit. Therefore, Thermodynamics and Classical Mechanics do not share the same conceptual structure and notion or \textit{ontology} \cite{batterman2002devil}.  It makes no sense to assign a temperature $T$ to a single particle since it is a macroscopic property of a bulk system, not of particles. Temperature is a \textit{local} and \textit{direct} emergent property \cite{carroll2024emergence}: local, because the temperature at any point depends only on particles within a nearby volume—not on distant parts of the system—thus preserving the spatial locality of the underlying theory; and direct, because the coarse-graining map is a simple analytic function—essentially the mean kinetic energy per particle, $T \propto \langle v^{2}\rangle$—rather than an algorithmically complex lookup.

Classical Mechanics itself may emerge from a quantum theory, which is, in turn, ontologically different \cite{landau2013quantum}. In Classical Mechanics, a particle is represented by a point in phase space with well-defined position and momentum. In Quantum Mechanics, the ontology shifts: a particle is represented by a vector in Hilbert space, and is described by a wavefunction \cite{griffiths2018introduction}. Position and momentum are no longer coordinates but observables. The uncertainty principle ensures that no wavefunction can define both position and momentum with perfect precision. Thus, quantum mechanics offers no definitive answer to the classical question of “where” a particle is, or “how fast” it is moving \cite{griffiths2018introduction}.
Nevertheless, the classic macroscopic theory remains applicable in its domain and is consistent with the underlying quantum description through decoherence and measurement, which effectively produce the appearance of wavefunction collapse \cite{zurek2003decoherence}.

\section*{Effective Theories}
Since the advent of Statistical Mechanics, we have learned how coarse-graining microscopic degrees of freedom gives rise to macroscopic quantities, such as temperature or pressure, and how these quantities are connected through equations of state \cite{kadanoff2009statistical}. 
Statistical mechanics not only explains why thermodynamics works but also defines its domain of applicability. For example, the domain of applicability of temperature can be expressed as
\begin{equation}
    T \;=\; \frac{1}{3}m\,\langle v^{2}\rangle\!\left(1 \pm \frac{1}{\sqrt{N}}\right),
\end{equation}
where $m$ is the particle mass, $\langle v^{2}\rangle$ is the ensemble-averaged squared speed, $N$ is the number of particles, and the Boltzmann constant is set to one, $k_{B}=1$ \cite{sethna2021statistical}. For large $N$, the fluctuation term $1/\sqrt{N}$ vanishes, making temperature a sharply defined coarse-grained variable. For small systems, however, fluctuations are significant, and temperature loses descriptive power. 

In the large-size limit, Thermodynamics is not merely an approximation to Statistical Mechanics—it is an \textit{effective theory} \cite{burgess2007introduction}, one that captures the essential behavior of macroscopic systems using a few key variables at the right scale \cite{wells2012effective, bedroya2024tale}.  Effective theories isolate what matters at a given scale and discard what does not \cite{shannon1948}. They are not shortcuts; they are self-contained, predictive, and often universal descriptions of emergent levels of reality. That is why, for specific questions, instead of tracking the detailed positions and velocities of every particle, we use variables like temperature, pressure, and entropy with their own simple and powerful macro-level laws \cite{bianconi2023emergence, shalizi2025macrostate}.
A \textit{mean-field theory} is the zeroth-order effective
description \cite{kivelson2024statistical}. It is invaluable as a first cut, exact when fluctuations
vanish, but insufficient whenever correlations hold the key to collective behavior \cite{porter2016dynamical}.

Not all effective theories are created equal. In some systems—particularly those involving biological, cognitive, or social dynamics—the coarse-grained variables do not merely summarize the current state of the underlying components; they also retain \textit{memory}, storing information about past configurations. Such memory and nonlocality are \emph{not} exclusive to living or social systems: in physics, integrating out fast fields generically produces spatially nonlocal interactions and temporal memory kernels  \cite{peskin2018introduction, feynman2000theory, zwanzig1961memory}. Even gravity has been cast in explicitly nonlocal, memory-like form \cite{mashhoon2014nonlocal, hehl2009nonlocal, hehl2009formal}. This historical dependence alters the structure of the effective theory, introducing higher-order terms that reflect feedback, path dependence, and self-reference \cite{carroll2024emergence}. Such systems are still governed by physical laws, but their dynamics can become computationally intractable, meaning that no tractable microscopic derivation will fully recover the macroscopic behavior \cite{dedeo2018origin}.

\section*{Onset of Emergence, Symmetry Breaking \& Criticality}\label{sec:sym}
A clear, intuitive example of emergence is magnetization. In a fridge magnet, the collective alignment of billions of electron spins (little magnets) produces a macroscopic magnetic property, denoted $m(T)$, which appears only when the system is below a critical temperature $T_c$. For a fridge magnet, it is around $450^{\circ}$C. Heating the material beyond this point destroys the alignment, and as Fig.~\ref{fig:coarse_graining_schem}d shows, the magnetic property disappears in a \textit{continuous phase transition} \cite{sethna2021statistical}. In other words, heating injects fluctuations that destroy spin alignment, driving the system from a magnetically ordered phase ($m>0$) to a disordered paramagnetic phase ($m=0$).
 A single governing equation (Ising Hamiltonian) describes both magnetized and non-magnetized states \cite{sethna2021statistical}.
 
 At the critical point $T = T_c$, the spin-flip symmetry of the Hamiltonian remains intact, but the actual state chooses a specific direction, resulting in $m \neq 0$ \cite{kivelson2024statistical, kulske2025ising}. This \textit{symmetry breaking} defines an \textit{order parameter} that compresses the full $10^{23}$-spin microstate into a single coarse-grained vector whose dynamics—like domain walls or spin waves—obey new effective laws \cite{peierls1936ising, griffiths1964ising, anderson1972more}.  
At  $T_c$, key properties of the system—such as correlation length and relaxation time—diverge  (or peak in finite systems), with finite-size scaling and data collapse as empirical checks  \cite{kadanoff1966scaling, fisher1972scaling, sethna2021statistical}. These divergences, characterized by critical exponents, serve as clear markers of the onset of emergence.

More interestingly, different systems composed of distinct elements can exhibit the same critical behavior \cite{griffiths1970dependence}. That is, systems with very different microstructures can fall into the same \textit{universality class} and be described by the same macroscopic theory. This \textit{substrate-independence} also means that, from the perspective of the emergent behavior, one cannot tell which microscopic system 
produced it.  When very different microscopic models display the same large-scale behavior, even exceptionally precise measurements of large-scale behavior usually cannot identify a unique microscopic theory without extra assumptions (e.g., symmetry content or locality) \cite{wilson1974renormalization,kadanoff1966scaling}.

The liquid–gas transition belongs to the same universality class as the ferromagnetic transition \cite{kardar2007statistical}. Fig.~\ref{fig:coarse_graining_schem}c illustrates the technique of translating these two transitions into one another. Universality classes, therefore, organize not just critical exponents but also the very symmetry content that survives at macroscopic scales, reinforcing why broken symmetry is one maker of emergence \cite{anderson1972more, witten2018symmetry}. In this sense, “more is different” \cite{anderson1972more}: once many degrees of freedom lock together, they generate effective laws—and sometimes entirely new effective symmetries—that are absent from, yet fully compatible with, the underlying dynamics.

It is worth mentioning that information-theoretic diagnostics similarly pinpoint the onset of emergence: mutual information rises near criticality, transfer entropy reveals increasingly predictive interactions, and persistent mutual information isolates long-lived choices beyond short-time correlations \cite{shannon1948,schreiber2000measuring,ball2010quantifying}. Causal tests complement these signals: Effective Information can increase under coarse-graining—so a well-chosen macro description exerts greater causal constraint than the micro—and partial information decomposition exposes downward causation and causal decoupling directly in multivariate data \cite{hoel2013quantifying,rosas2020reconciling}.

\section*{Renormalization \& Universality}
Technically, critical universality manifests only within the \emph{critical region} of a continuous phase transition: as control parameters approach the transition, the \emph{renormalization-group} (RG) flow is drawn toward a scale-invariant fixed point—attractive along irrelevant directions and unstable along relevant ones—so macroscopic behavior becomes largely independent of microscopic details \cite{wilson1974renormalization, binney1992theory, kupiainen2023rigorous}. This picture applies to both equilibrium and non-equilibrium \emph{continuous} transitions \cite{henkel2008non,hinrichsen2000non}; nothing forbids far-from-equilibrium universality \cite{tu2023renormalization}, though for certain many-body Hamiltonians even basic features (spectral gap, phase diagram, etc.) are \emph{undecidable in principle} despite a computable RG flow \cite{watson2022uncomputably}.

 A canonical non-equilibrium exemplar is the class of \emph{absorbing-state} transitions \cite{ takeuchi2009experimental, badie2022directed, badie2022directed2}: dynamics are irreversible, the system can enter configurations it cannot escape, and the order parameter (e.g., the active-site density in the contact process) vanishes continuously at a critical control parameter with power-law correlations at criticality \cite{henkel2008non, hinrichsen2000non}. Under broad conditions, these phase transitions fall into the \emph{directed percolation} (DP) universality class, often regarded as the non-equilibrium analogue of the Ising paradigm we mentioned before \cite{henkel2008non, hinrichsen2000non}. 
 Neural-population recordings, deep neural networks, and insect swarms also exhibit RG-style scaling in nonequilibrium settings \cite{friedman2012universal,tkavcik2015thermodynamics,meshulam2019coarse,ghavasieh2025toward,cavagna2017dynamic, cavagna2023natural}.
What truly matters for RG are locality, symmetry, and conservation constraints, and dimensionality \cite{henkel2008non, tauber2014critical}. That said, beyond a few well-understood universality classes, we still lack a clear, well-founded theoretical framework—comparable in scope to the renormalization group—for systematically organizing and classifying \emph{non-equilibrium} phases \cite{henkel2008non}.
 
One must take care to remember that criticality is neither the only route to emergence nor trivial to establish empirically \cite{cross1993pattern}.
 For example, it is not straightforward to extend the statistical mechanics framework of critical phenomena to describe abrupt qualitative changes in dynamical systems. Likewise, the emergence of life from chemical interactions does not neatly correspond to a phase transition in the physical sense, despite claims to the contrary by some physicists \cite{bak2013nature}. Intricate phenomena such as consciousness or thought could, in principle, arise from entirely different physical substrates—biological neurons or silicon circuits alike \cite{oizumi2014phenomenology, tononi2016iit}—provided there is sufficient evidence that they result from well-defined \textit{critical phenomena} \cite{berche2009fenomenos}. Across biological systems, current evidence favors near-criticality over exact self-tuned criticality \cite{beggs2012being, o2022critical, munoz2018colloquium}. In neuroscience in particular, recent surveys argue that a distance-to-criticality perspective is more realistic and useful—for cognition and pathology—than the claim that the brain sits precisely at criticality; both edge-of-chaos and avalanche-like regimes can appear, often state-dependently \cite{beggs2012being, o2022critical, munoz2018colloquium}. Therefore, “critical-looking” statistics (power laws, avalanches) are insufficient on their own \cite{clauset2009power, friedman2012universal, stumpf2012critical, touboul2010can, neto2022sampling} and broken symmetries underpin not all emergent behaviors \cite{strogatz2022more,vafa2020puzzles}. 
Accordingly, strong claims of criticality, let alone emergence in living systems, require careful analysis \cite{fisher1972scaling, dakos2012methods, sethna2021statistical, henkel2008non}. 

\section*{Duality \& Emergence}
Duality is an equivalence (often infrared-exact \cite{de2025philosophy, castellani2020duality}) between two seemingly different theories: a one-to-one “dictionary” that maps observables, states, and parameters so that all physical predictions coincide. Dualities reveal that theories with ontologically distinct variables—such as bosons versus fermions, or gravity versus no gravity—can describe the same underlying physics \cite{de2025philosophy}. 
Formally, a duality establishes a structural correspondence between the measurable quantities of two theories (a *-isomorphism) such that every observable, expectation value, and response function matches once external conditions—sometimes called “sources”—are translated accordingly. This agreement may accommodate local adjustments, such as contact terms or counterterms, reflecting different renormalization schemes \cite{de2025philosophy}.

In \(1{+}1\) dimensions, the bosonic sine–Gordon model is exactly dual to the fermionic Thirring model, exchanging solitonic waves for interacting particles and relating their couplings nonlinearly \cite{coleman1975quantum}. In the Anti–de Sitter/conformal field theory correspondence, a conformal field theory without gravity on a \(d\)-dimensional boundary is holographically equivalent to a \((d{+}1)\)-dimensional bulk theory \emph{with} gravity \cite{maldacena1999large}. Classic examples from statistical and gauge theories—such as high–low temperature mappings, electric–magnetic duality, and particle–vortex correspondence—further illustrate how distinct microscopic descriptions can encode the same macroscopic behavior \cite{savit1980duality, montonen1977magnetic, seiberg1994monopoles, peskin1978mandelstam}

What we learn from dualities is twofold: First, large-scale behavior is not dictated by microscopic “stuff” (but by \textit{invariants} preserved under the duality map—correlators, partition functions, anomalies, and consistency with locality and boundary conditions) \cite{castellani2020duality}. Second, because dualities often relate weak and strong coupling, they reveal multiple, equally valid languages for describing the same phenomena \cite{castellani2020duality}. Macroscopic structure, therefore, can appear as alternative readouts of a single physical substrate. Which language proves most useful depends on the question at hand—computational tractability, available probes, or symmetry constraints—rather than on a uniquely privileged ontology.

\begin{figure*}[hbt]
    \centering
\includegraphics[width=.89\linewidth]{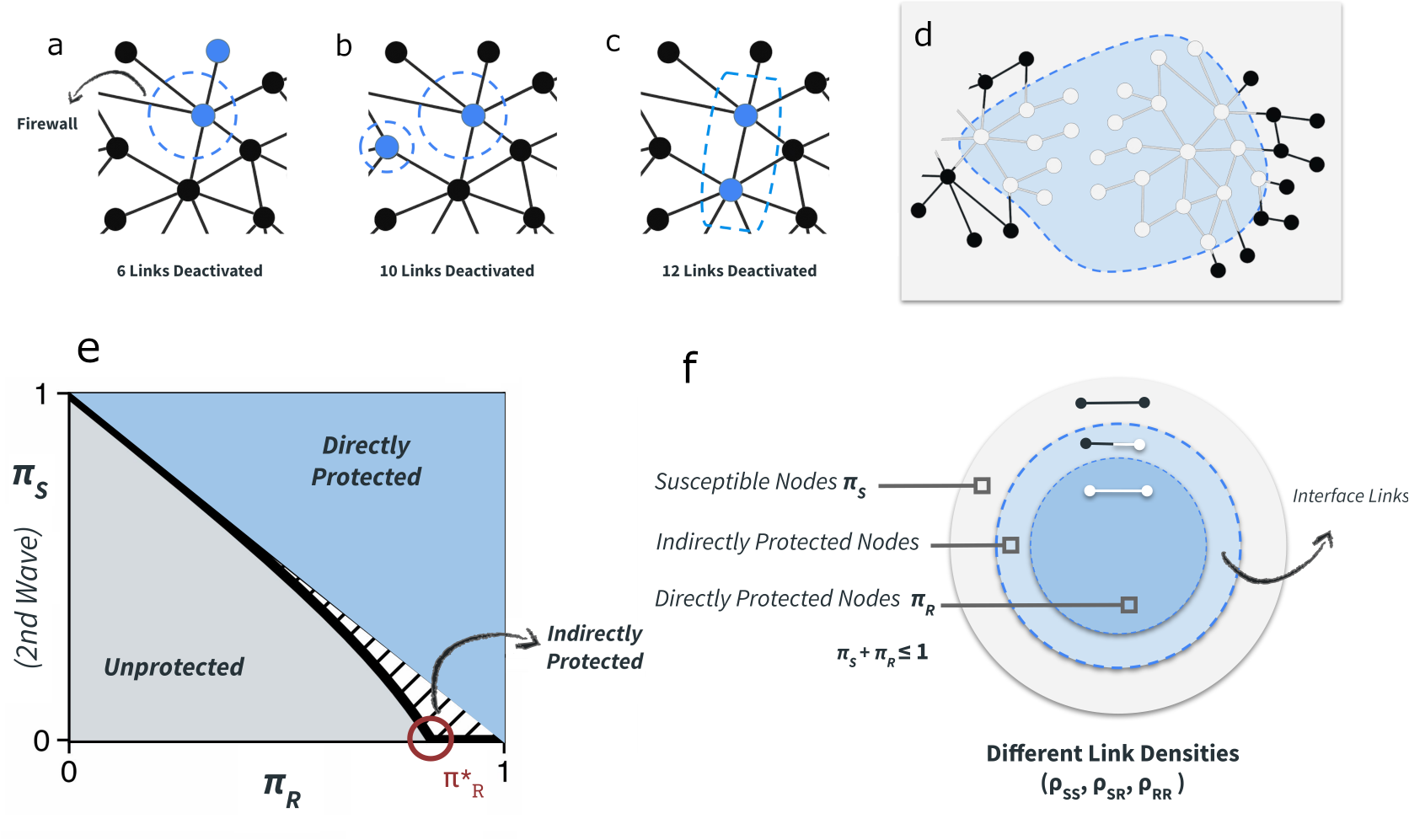}
    \caption{
\textbf{Emergence of herd immunity in social networks.} 
\textbf{(a–c)} Each immune node (blue) deactivates its adjacent edges, forming a local “firewall” that halts transmission to nearby susceptible nodes. The more connections an immune node has, the larger the firewall it establishes. As immunization progresses, these firewalls begin to overlap and coalesce, forming a larger collective barrier that expands nonlinearly—often outpacing the fraction of immunized individuals.  
\textbf{(d)} Once a critical fraction of nodes (white) is immunized, herd immunity emerges: the remaining susceptible nodes (black) become indirectly protected. This emergent protection is shaped by both the structural and geometric properties of the social network.  
\textbf{(e)} The thick solid curve shows the remaining susceptible fraction \(\pi_\mathcal{S}\) as a function of the immunized fraction \(\pi_\mathcal{R}\), constrained by \(\pi_\mathcal{S} + \pi_\mathcal{R} \leq 1\). The shaded region quantifies the share of individuals shielded through structural (indirect) immunity. Horizontal intersections of the curve indicate total immunity thresholds, with \(\pi^*_\mathcal{R}\) marking the structural herd immunity point.  
\textbf{(f)} The network in panel (d) can be rearranged to reveal an interface of \(\mathcal{SR}\) links separating immune (white) from susceptible (black) nodes. The density of these interface links, \(\rho_{\mathcal{SR}}\), serves as a proxy for the potential of epidemic containment and the strength of indirect protection.
}
    \label{fig:enter-label}
\end{figure*}

\section*{Emergence in Networks \& Social Systems}
Moving beyond physics, we consider how large-scale patterns emerge from local interactions in social and informational systems. Emergent phenomena possess causal power. In both the physical and social worlds, entities such as society and culture act as real forces shaping outcomes. Even in systems designed without formal rules—such as decentralized communities like Wikipedia or Stack Exchange—structures, hierarchies, and power distributions inevitably emerge, shaping the world in profound ways.

Consider a set of $N$ points where each pair is connected with probability $p$. This defines an Erdős–Rényi random graph with $N$ \textit{nodes} and roughly $p\binom{N}{2}$ \textit{links} or \textit{connections}. As $p$ increases, the network shifts from isolated nodes ($p = 0$) to a fully connected graph ($p = 1$), with clusters forming in between. Given the value of $p$, it is interesting to look at the behavior of the cluster with the largest number of connected nodes, the \textit{giant component}. Letting $\sigma$ be the fraction of nodes it contains, we find that as in the large-size limit ($N \to \infty$), the network undergoes a phase transition such that the size of its giant component is zero below a critical value $p_c$, and positive above it.
Here, $p$ is the control parameter, and $\sigma$ represents the order parameter, mirroring magnetization in ferromagnets, which appears below the critical temperature. When $p > p_c$, large-scale connectivity emerges, and the network is said to percolate \cite{newman2018networks}.

The emergence of the giant component is foundational: the Internet stays functional despite random failures because its giant component remains intact \cite{newman2018networks}. Likewise, in epidemic dynamics, if an infected person enters a social network and contacts someone in the giant component, the disease can potentially spread to nearly everyone after enough time. In this way, the onset of an epidemic reflects the same percolation principles that govern network connectivity \cite{newman2002spread, k2024spreading}. We say that an outbreak occurs or an epidemic emerges when a significant number of individuals in a population become infected. The onset, size, and time scales of this emergence can be measured and modeled by examining the disease's contagiousness and the structure and changes in the contact network \cite{pastor2015epidemic}. The emergence of large-scale connectivity is one example among many in networks. Furthermore, it is also easy to argue that communities emerge when networks evolve with local rules \cite{Vazquez}.

A mesoscale theory can clarify what aspects a good coarse-graining should focus on \cite{mckenzie2025emergencephysicsbiologysociology}. For instance, when describing connectivity and spreading phenomena on networks at an intermediate level—specifically, how communities interact—it can capture most of the predictive signals while disregarding unnecessary details \cite{rizi2024effectiveness}. At this scale, the dynamics are largely self-contained, with percolation and epidemic thresholds determined mainly by the pattern and strength of inter-module links \cite{gabrielli2025network}. Identifying this mesoscale is an integral part of the theory itself \cite{alessandretti2020scales, edsberg2022understanding}, as it is where compression, autonomy, and universality come together \cite{mckenzie2025emergencephysicsbiologysociology}.
\section*{Emergence of Herd Immunity in a Population}
Vaccination is a key strategy for suppressing epidemics. Take measles, a highly contagious disease with a basic reproduction number of $R_0 \sim 15$—meaning one case can infect around fifteen others early on in a fully susceptible population \cite{anderson1991infectious, guerra2017basic}. Before routine vaccination in the UK began in the late 1960s, outbreaks recurred every 2–3 years as the number of susceptible children quietly rose above the threshold for transmission.

Vaccination protects in two ways: \textit{directly}, by immunizing individuals and removing them as nodes from the transmission network, and \textit{indirectly}, as Fig.~\ref{fig:enter-label}(a,b) shows, by forming a \textit{firewall} (or \textit{firebreak} \cite{branda2025computational}) and breaking the chains along which the infection might spread \cite{hiraoka2022herd}. In a vaccinated population, before the introduction of a new infection, individuals are either susceptible ($\mathcal{S}$) or immune ($\mathcal{R}$). The network representation of this population can always be rearranged to separate these two groups, with the interface formed by links connecting $\mathcal{S}$ and $\mathcal{R}$ nodes (Fig.~\ref{fig:enter-label}f). The size of the immune set represents the direct benefit of vaccination, while the number of interface links serves as a proxy for indirect protection.

When enough immune individuals form clusters, they fragment the transmission network into disconnected parts (Fig.~\ref{fig:enter-label}d), stopping the infection from spreading widely. In this way, immunity percolates through the population: local outbreaks fail to propagate beyond their immediate neighborhood, and as a result, even unvaccinated individuals benefit from this collective protection, known as \textit{herd immunity} \cite{newman2018networks}. The extent of the emergent barrier—the \emph{firefront}—is quantified by the density of susceptible-immune ($\mathcal{SR}$) links, $\rho_{\mathcal{SR}}$. Each $\mathcal{SR}$ link both blocks transmission and reduces the downstream branching factor of the pathogen, making $\rho_{\mathcal{SR}}$ a nonlinear, structural measure of resistance. The question, then, is not just \emph{how many} people must be vaccinated, but \emph{who} should be vaccinated \textit{with what order} for herd immunity to emerge?

In a hypothetical fully mixed population---where each individual interacts with all others with equal probability---this question has a simple answer. If a fraction \(\pi_\mathcal{R}\) of the population is randomly vaccinated before the introduction of the disease, then a typical infected person can effectively infect $R_e = (1 - \pi_\mathcal{R}) R_0$ more people.
Outbreaks become unsustainable when \( R_e < 1 \), yielding the classical threshold. $
\pi_\mathcal{R}^* = 1 - \frac{1}{R_0}.$
This formulation, however, assumes that immunity acts uniformly and independently across individuals, neglecting the underlying contact structure of real populations \cite{hiraoka2022herd}.
However, real populations form heterogeneous, spatially embedded networks \cite{ghadiri2024impact}. Most interactions cluster within social, geographic, or institutional contexts \cite{rizi2024homophily}.

A node’s number of connections is referred to as its degree \cite{newman2018networks}. Immunizing a highly connected node can disrupt many transmission routes, whereas vaccinating randomly chosen low-degree nodes typically offers little indirect protection. Yet, the combined impact of removing multiple nodes is not easy to predict: the strength of the resulting indirect immunity depends sensitively on the network’s spatial structure and the distribution of immunized nodes within it \cite{hiraoka2023strength}.
As Fig.~\ref{fig:enter-label}(a-c) shows, each immune node acts as a local firewall, suppressing links through which infection would otherwise spread. As immunity accumulates, these firewalls overlap, and their collective effect can grow faster than the immunized fraction (Fig.~\ref{fig:enter-label}d). 
By immunizing a fraction $\pi_\mathcal{R}$ of the population, at most a fraction $\pi_\mathcal{S} \leq 1 - \pi_\mathcal{R}$ remains susceptible. The exact value of $\pi_\mathcal{S}$ depends on the network's structure and how immunity fragments it. The difference, $1 - \pi_\mathcal{R} - \pi_\mathcal{S}$, corresponds to individuals who are not vaccinated but are indirectly protected, represented by the hatched area in Fig.~\ref{fig:enter-label}e.  

Recent studies formalize this mechanism using bond percolation and message-passing on real and synthetic networks \cite{hiraoka2023strength, ghadiri2024impact}. They track the dynamics of the susceptible giant component and the firefront density, $\rho_{\mathcal{SR}}$, across diverse mixing patterns. These analyses reveal a competing mechanism: targeting superspreaders or implementing acquaintance immunization increases collective immunity, whereas the network's spatial structure can hinder it by localizing protection, leaving remote, susceptible pockets vulnerable \cite{hiraoka2023strength}. 

Herd immunity is a paradigmatic case of emergence in social networks: local immunizations combine nonlinearly to produce a global shield shaped by network structure. Recognizing this deepens our grasp of collective behavior and offers practical tools for designing more efficient interventions. Yet translating this insight into policy is non-trivial, given heterogeneous and mobile populations, vaccine profiles that differentially affect severity, infection, and carriage, and constraints on supply, waning, immune escape, uptake, equity, and ethics. Network-informed models should therefore complement—rather than replace—clinical priorities and socioeconomic evaluation, chiefly to enhance indirect protection where appropriate. The aim is disciplined decision support, not a prescriptive blueprint.

\section*{Weak vs.\ Strong Emergence}
From the standpoint of science, there is no mystery or divinity in emergence. If a property is observable (measurable), it is physical, and any scientific explanation must reference the constituents and their interactions, no matter how surprising the outcome appears.  Nevertheless, the consistency between theories with different ontologies raises the question of how much \textit{causal autonomy} higher levels truly possess—a distinction philosophers frame as weak versus strong emergence.
In science, emergence refers to collective behaviors that have explanatory autonomy \cite{fodor1974special} and are, in principle, possible but practically difficult to derive from purely microscopic descriptions, even when all micro-level details are known \cite{oconnor1994}. Philosophers often refer to this as \textit{weak emergence} \cite{bedau1997}, while leaving room for the idea of \textit{strong emergence}—where some phenomena might be fundamentally irreducible and governed by distinct principles \cite{chalmers2006strong}. 
Strong emergence suggests that nature might include layers that follow their own fundamental rules—possibly violating physical laws, creating inconsistencies between micro and macro theories, or breaking causal closure or locality.

 It is worth mentioning, that one may argue that some microscopic theories are incomplete or inapplicable to specific macroscopic regimes—but to invoke strong emergence in domains like consciousness or social behavior \cite{kim2000mind, baronchelli2018emergence} amounts to introducing new causal principles not anchored in substrate dynamics, and thus steps outside the scope of \textit{scientific method} \cite{gauch2003scientific}. Therefore, if a macro-level description \textit{effectively} explains a phenomenon—even more \textit{parsimoniously}, with fewer micro-level details than a microscopic one—nothing beyond \textit{physicalism} is at play. There have been, of course, historical movements that made strong-sounding claims about emergent phenomena without yielding testable or predictive frameworks \cite{kirchner2002gaia, bak1987self, stumpf2012critical, maturana1972autopoiesis, bich2024autopoiesis, bedau2000open}.

As causality connects events within the same scale, emergence connects descriptions across scales, revealing how collective patterns arise from local interactions. But the influence does not only run upward. Much like the walls of a container shape the motion of the particles inside, higher-level structures can constrain or guide the behavior of their components without violating the underlying laws \cite{hobson2015social}. This interplay is often described as downward or top-down causation \cite{campbell1974, ellis2008top, flack2017coarse}. Dominance hierarchies in animal groups illustrate this bidirectional linkage: dyadic encounters build rank, and that rank feeds back to channel individual aggression toward nearby rivals \cite{hobson2015social}.

It is tempting to wonder whether complex aesthetic outcomes might also emerge from simple, knowable constraints—a question deeply resonant with the concept of emergence itself. If Poe’s self-portrait in \textit{The Philosophy of Composition} (1846) is taken at face value, at least one kind of aesthetic complexity is not \textit{strongly} emergent; it can be reconstructed from an explicit, step-wise recipe. Poe famously crafted the poem \textit{The Raven} by first fixing a desired emotional effect—melancholy—and then deliberately selecting every element (theme, meter, and the refrain “Nevermore”) to achieve that goal \cite{poe1924philosophy}. He insists good writing is methodical and analytical, explicitly rejecting artistic intuition. While many, including T. S. Eliot, argue that something ineffable lies beyond algorithmic methods \cite{eliot1992criticize}, Poe’s essay reminds us that not all creativity is opaque to analysis—even if many artists and readers prefer it that way. Today, sentiment-analysis tools can map—and perhaps replicate—a narrative’s emotional arc \cite{reagan2016emotional, pan2025story}, sharpening our insight into how writers engineer intricate effects. However, until we can define such processes rigorously and run controlled experiments on well-defined creative variables, any final judgment about whether art is ultimately rule-bound or irreducibly spontaneous must remain tentative.

\section*{Conclusion and Discussion}
There is a story in Persian about a modest art instructor who could skillfully draw many animals—rabbits, deer, birds—but always avoided one: the horse. One day, the students insisted that he draw a horse. Reluctantly, he began from the head, moved gracefully down the body, but as he reached the legs and hooves—his weak point—he hesitated. Then, with a swift stroke, he drew tall grass over the lower legs, hiding the part he could not render. When the students asked him why he added grass, he replied that horses naturally belong in fields, neatly sidestepping the truth. In many scientific papers, the term emergence is treated as if it were with the hooves. The arguments are precise and formal until they reach a point that resists proper modeling or measurement, at which point people resort to vague arguments to cover it up. Lacking consensus on how to define or observe emergence as a physical quantity, authors often cloak conceptual gaps with terms that sound profound but remain undefined. Such ambiguity risks obscuring what is truly being explained. 
 
 In this work, we show the onset of emergence can be located through complementary lenses: in symmetry-breaking transitions, an order parameter turns on at a critical point while susceptibility and correlation length diverge; in RG terms, tuning the control parameter drives the flow to a scale-invariant fixed point so macroscopic behavior becomes largely insensitive to microscopic detail \cite{wilson1974renormalization,binney1992theory,kupiainen2023rigorous}. We seek to clarify emergence by framing it as a many-to-one coarse-graining process that retains predictive power while discarding most micro-level details. Within this framework, we differentiate between the [weak] emergence, which is fully compatible with underlying dynamics, and the more philosophical notion of strong emergence. Finally, as a concrete example in a more social setting, we mentioned recent work on herd immunity in structured contact networks where emergence can be rigorously defined, quantified, and linked to network geometry. Understanding the emergence of herd immunity highlights that epidemic control, like many collective phenomena, cannot be fully understood without considering the structure and correlation across social scales.

 Emergence is not a license to skip reason, nor a metaphysical extra cause \cite{bedau2008emergence}. Higher-level patterns (from thermodynamic properties to life itself) can be real and explanatory while remaining grounded in physical laws \cite{dennett1991real}. Any genuine emergent feature must still emerge from somewhere, meaning there is always an underlying mechanism, not beyond the standard physical interactions, producing it.\cite{de2019towards}. Science is compatible with a form of pluralism that affirms the reality of higher-level causal powers \cite{simpson2022toppling}. As scientists, we must use language carefully and remain grounded in physical mechanisms. The word 'emergence' is widely used in the field of Complex Systems. While vague or mystical references to emergence may sound compelling, selling complexity science by mystifying emergence and invoking some spiritual dimensions is a great disservice \cite{petterBlog2025}.

\acknow{AKR thanks Sune Lehmann, Petter Holme, Yasser Roudi, Hossein Mahdaei, and Behrad Taghavi for insightful discussions related to this work.
This project was supported by the Independent Research Fund Denmark (EliteForsk grant to Sune Lehmann), the Carlsberg Foundation (The Hope Project), and the Villum Foundation (NationScale Social Networks).
AKR declares no competing interests.
Language and style were refined with the assistance of AI tools, under the author’s full supervision and final approval. An earlier version of this manuscript was posted as a preprint \cite{rizi2025emergence}. This work does not involve underlying data.

}

\showacknow{} 
\bibliography{citations}
\end{document}